\documentclass[conference]{IEEEtran}
\IEEEoverridecommandlockouts

\usepackage[shortcuts]{extdash}

\usepackage{graphicx}
\usepackage{subcaption}
\usepackage{multirow}
\usepackage{siunitx}
\usepackage{cleveref}
\crefname{section}{Section}{Sections}
\crefname{appendix}{Section}{Sections}
\crefname{figure}{Figure}{Figures}
\Crefname{figure}{Figure}{Figures}
\crefname{table}{Table}{Tables}
\crefname{equation}{Eq.}{Eq.}
\Crefname{equation}{Equation}{Equations}

\usepackage{soul}

\usepackage{xspace}
\ifCLASSOPTIONcompsoc
  \usepackage[nocompress]{cite}
\else
  \usepackage{cite}
\fi

\usepackage{fancyhdr}

\ifCLASSINFOpdf
\else
\fi

\begin{document}

\title{Supporting Dynamic Control-Flow Execution for Runtime Reconfigurable Processors\\

\thanks{This work was partially funded by the ``Helmholtz Pilot Program for Core Informatics (kikit)'' at Karlsruhe Institute of Technology. The authors would like to thank Niklas Lorenz, Sascha Herring, and Andreas Bühner for their help in implementing the accelerators.}}

\author{\IEEEauthorblockN{Hassan Nassar, Rafik Youssef, Lars Bauer, Jörg Henkel}\\
\IEEEauthorblockA{\textit{Chair for Embedded Systems, Karlsruhe Institute of Technology, Germany}\\
hassan.nassar@kit.edu}
}

\IEEEtitleabstractindextext{%
\begin{abstract}
As the need for more computing power grows, traditional methods are hitting limits. 
To boost performance, we're expanding Central Processing Unit (CPU) capabilities and using specialized hardware accelerators.
For example, mobile devices usually have cameras, video encoding, and audio accelerators.
To perform the different tasks, these accelerators execute microcode programs.
These accelerators, however, take up space and often sit idle. 
Reconfigurable processors offer a solution. They have a normal core connected to several accelerator slots.
These accelerator slots can be filled during runtime to accommodate the application running.
Once one application finishes and another application is running, the accelerators can be switched.
For example, playing music after using the camera.

In this work, we introduce dynamic control-flow execution for the microcode of runtime reconfigurable processors, i.e., support for loops, conditional jumps, and exception handling.
We benchmark using four different applications from four domains (object detection, ocean movement simulation, artificial intelligence and security) that all are compute-intensive and would require the dynamic control-flow when executed on reconfigurable processors.
We show that the dynamic control-flow allows different applications to be executed with significant speedup in comparison with execution on general-purpose processors.
\end{abstract}

}

% make the title area
\maketitle
\renewcommand{\headrulewidth}{0.0pt}
\thispagestyle{fancy}
\lhead{}
\rhead{}
\chead{This is the authors' preprint version of the work.
The definitive Version of Record is published in the 2023 IEEE International Conference on Microelectronics (ICM).}
\cfoot{}
\IEEEdisplaynontitleabstractindextext

\IEEEpeerreviewmaketitle

\section{Introduction}

The conventional methods for enhancing processors have hit their technological limits, necessitating fresh approaches to meet the increasing demands of applications. 
One such approach is the Application-Specific Instruction Set Processor (ASIP), which enables a higher level of parallelism. 
These processors are fairly basic, featuring a reduced instruction set architecture (RISC). 
There is also the option to introduce instruction set extensions or incorporate special instructions (SIs) to enhance their functionality. These SIs, much like Application-Specific Integrated Circuits (ASICs), boast unique hardware implementations that enable them to achieve greater parallelism~\cite{bauer2009rispp}.

Field Programmable Gate Arrays (FPGAs) offer a very attractive solution for hardware acceleration.
FPGAs consist basically of Look Up Tables (LUTs) which are able to implement basic logic gates.
By interconnecting the LUTs and usage of registers they can implement anything from simple logic circuits up to processors and even more complex hardware. 
Therefore, FPGAs are used in a wide range of applications, e.g., digital signal processing, cloud computing, and machine learning acceleration~\cite{iccad21,icm22,muteco21,NGCAS2017DPRNOC}. 

By combining CPUs performing SIs with FPGAs and employing reconfiguration or alternative configuration streams, the hardware implementation can be modified, potentially even during runtime~\cite{bauer_fau2022}. 
This capability paves the way for the exploration of novel strategies and facilitates the development of reconfigurable systems.
This is especially relevant for resource-constraint devices, e.g., mobile devices where the resource allocation can be challenging.
Runtime reconfiguration gives the chance to load the accelerators only when needed and to switch from one accelerator to another whenever needed.

\begin{figure}[!t]
    \centering
    \includegraphics[clip=true, width=1\linewidth,page=1]{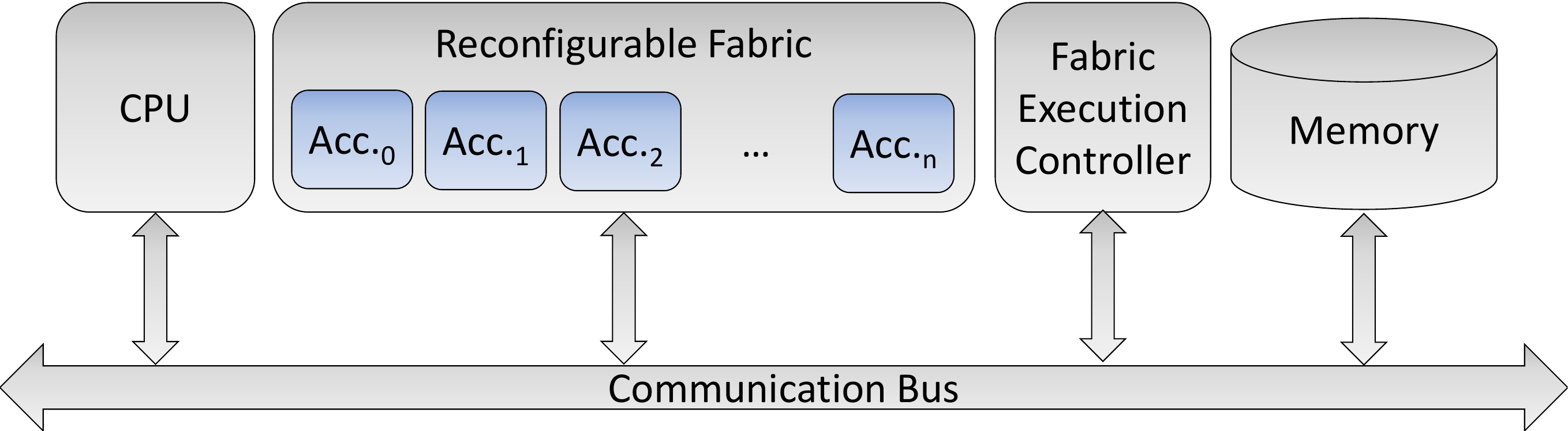}
    \caption{Target Reconfigurable Processor Architecture. A main CPU core is connected to a reconfigurable Fabric. The execution of the accelerators residing on the fabric is controlled by a controller executing microcode.}
    \label{fig:sys_overview}
    \vspace{-2em}
\end{figure}

A typical reconfigurable processor would look as shown in \cref{fig:sys_overview}.
A CPU is connected to a Reconfigurable Fabric which in itself is divided into several accelerator slots.
Each accelerator slot can be used to load the different accelerators at runtime.
The accelerators execute the SIs which consist of microcode performed by the accelerator.
A fabric execution controller controls the execution of the SIs~\cite{bauer_fau2022,bauer2009rispp}. 

Applications accelerated on such systems can be rather complex.
Therefore, the microcode executed needs to be able to fulfill all needed control steps of the application.
Such applications might need jumps, conditional loops that cannot be unrolled, etc.
Therefore, in this work, we tackle the issue of having a dynamic control flow for reconfigurable processors.
Our novel contributions are as follows:
\begin{itemize}
    \item We introduce dynamic control flow execution support to  accelerators executing SIs of reconfigurable Processors
    \item We analyze several applications on how to make them benefit from the dynamic control flow execution
    \item We show the timing improvement in executing these applications on hardware utilizing the dynamic control flow
\end{itemize}

The rest of the paper continues as follows.
The background needed is given in \cref{sec:back}.
We present our implementation for the dynamic execution in \cref{sec:method}.
The benchmark applications are detailed in \cref{sec:bench}.
We evaluate our results in \cref{sec:results} and draw conclusions in \cref{sec:conclusion}.

%%%%%%%%%%%%%%%%%%%%%%%%%%%%%%%%%%%%%%%%%%%%%%%%%%%%%%%%%%%%%%%%%%%%%%%%%%%%%%%
%\clearpage
\section{Background}
\label{sec:back}

\begin{table*}[!t]
  \centering
  \caption{Dynamic Execution jump sub-instructions of the SI VLIW. The different jumps allow it to go dynamically within the microcode based on accelerator output or user-defined counter values. Hence, more complicated algorithms can be accelerated on the reconfigurable fabric.}
  \label{tab:SI_instr}
  \begin{tabular}{|c|c|c|}
    \hline
    Instruction & Arguments & Explanation \\
    \hline
    NO\_JMP & none & don't jump \\
    \hline
    ALW\_JMP & destination & jumps always to destination\\
    \hline
    JMP\_IF\_CNT\_EQ & destination, value & jump to destination if counter equals value\\
    \hline
    JMP\_IF\_CNT\_NEQ & destination, value & jump to destination if counter does not equal value\\
    \hline
    JMP\_IF\_CNT\_LT & destination, value & jump to destination if counter is less than value\\
    \hline
    JMP\_IF\_CNT\_GT & destination, value & jump to destination if counter is greater than value\\
    \hline
    JMP\_IF\_ACC\_EQ & destination, value, accelerator slot & jump to destination if output from accelerator equals value\\
    \hline
    JMP\_IF\_ACC\_NEQ & destination, value, accelerator slot & jump to destination if output from accelerator does not equal value\\
    \hline
    JMP\_IF\_ACC\_LT & destination, value, accelerator slot & jump to destination if output from accelerator is less than value\\
    \hline
    JMP\_IF\_ACC\_GT & destination, value, accelerator slot & jump to destination if output from accelerator is greater than value\\
    \hline
  \end{tabular}
\end{table*}

Reconfigurable Processors are continuously gaining attraction.
From academia, several architectures have been proposed, e.g., RISPP~\cite{bauer2009rispp}, \textit{i}-Core~\cite{bauer_fau2022}, Molen~\cite{Molen}, and KHARISMA~\cite{kharisma}.
From industry, the Zynq MPSOC from Xilinx-AMD combines embedded ARM processing system (PS) with a reconfigurable programmable logic (PL)~\cite{zynq}.
Similarly, Intel provides the Intel Stratix SoCs with similar architecture~\cite{stratix}.

The reconfigurable fabric of reconfigurable processors is usually divided into several accelerator slots as \cref{fig:sys_overview} shows.
The reconfigurable fabric can then be used by the applications running on the GPP to accelerate their hotspots and guarantee that they meet their timing requirements~\cite{iccad19}.
This is advantageous as for out-of-order processors which can also increase the performance, guarantees for meeting timing requirements are not provided~\cite{iccad19}.

The accelerators residing on the reconfigurable fabric execute SIs consisting of a microcode.
It consists of the successive steps that have to be executed for the accelerator to complete its intended task.
The microcode of the SI is usually given to the reconfigurable fabric in VLIWs~\cite{bauer_fau2022,bauer2009rispp,autoSI}.
Each VLIW consists of several sub-instructions with each sub-instruction controlling the execution on one of the accelerators.
The reconfigurable fabric can even be shared between multiple GPPs~\cite{corefab}.
If no conflict occurs, e.g., two applications requiring different sets of accelerators, the VLIWs for two or more SIs can be merged together and the performance on several cores will be enhanced.

When reconfigurable processors execute Special Instructions (SIs), the control flow is defined in a static manner. 
The VLIWs are processed sequentially, one after another, with no opportunity for the loaded accelerators to influence the program flow. 
At maximum, loops that can be unrolled will be unrolled and executed~\cite{autoSI}.
To address this constraint, we develop a dynamic execution control system for these special instructions. 
\textit{To the best of our knowledge, our work is the first to tackle the need and the implementation of dynamic execution of reconfigurable processors.}
%%%%%%%%%%%%%%%%%%%%%%%%%%%%%%%%%%%%%%%%%%%%%%%%%%%%%%%%%%%%%%%%%%%%%%%%%%%%%%%%%%%%%%%%%%%%%%%%%%%%%%
\section{Dynamic Control-Flow Support}
\label{sec:method}

Several features are needed to support the acceleration of more complex algorithms.
Firstly, there is a need to introduce support for jumps within the SIs. These jumps can function as conditional statements, enabling the creation of loops within the execution flow. 
Secondly, the inclusion of exception-handling mechanisms is essential. 
This is required not only to identify errors in the program flow but also to handle errors originating from the accelerators. 

To incorporate the necessary functions, it is imperative to introduce new commands into the existing VLIWs. 
Each VLIW serves as a map for commands related to all components within the fabric, including the control logic. 
All the parameters related to jumps, along with their predefined structure, are established in advance using distinct commands. 
To enable the implementation of concepts like nested loops within the SI, there are four separate sets of parameters that are internally reserved by the controller. 
Consequently, a jump command encapsulates the entire parameter set, requiring only 2 bits to represent it. 
Each of these parameter sets includes a counter spanning 12 bits in length. 
This choice balances command length and the range of values it can accommodate. 

Our implementation offers three distinct categories of jump commands. 
The first group comprises static commands that are executed unconditionally. 
Within this group, we have "No Jump," which signifies that no jump should occur. 
Notably, this command doesn't require a parameter set. 
Additionally, the static commands include "Always Jump," an instruction that unconditionally redirects execution to its specified destination.

The second category encompasses jumps whose conditions are tied to one of the four counters.
These jumps, in addition to the parameter set, require an operand that is compared with the counter to determine whether the jump occurs.
The final category comprises conditional jumps, where the conditions are associated with a 2-bit control signal from one or more accelerators. 
These jumps will only occur if the control signals of all selected accelerators meet the specified criteria. 
In addition to the parameter set and the operand, the selection of the accelerators is also required for these jumps.
It's important to note that all conditional jumps, whether they rely on the counter or control signals, are available in four distinct variants.
\Cref{tab:SI_instr} shows all the newly added jump instructions.

\begin{figure}[htbp]
    \centering
    \includegraphics[clip=true, width=1\linewidth,page=1]{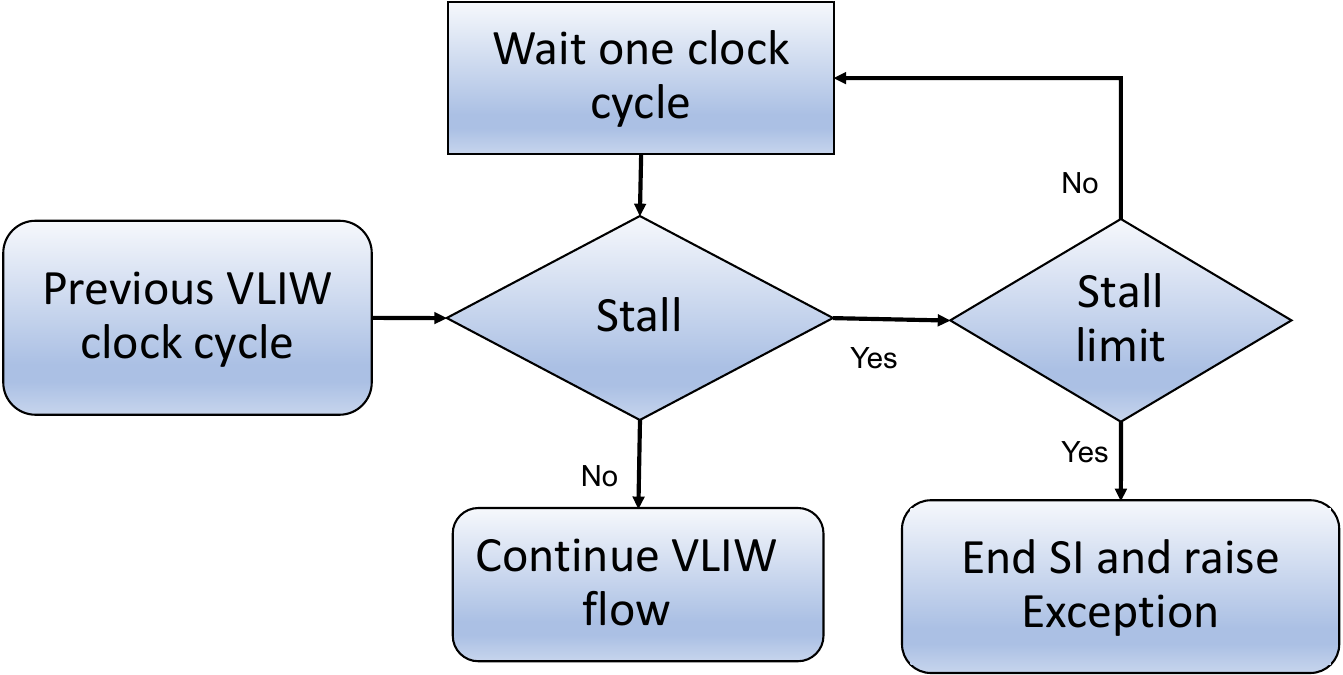}
    \caption{Exception Support for the Dynamic Control-Flow}
    \label{fig:stall}
    \vspace{-2em}
\end{figure}

We assume that the underlying CPU comes equipped with exception-handling capabilities. 
Therefore, we use this to create our support for exceptions.
The first group of traps is logic related. 
An invalid jump target, that falls outside its permissible boundaries triggers a trap event.
If one of the accelerators runs into an error, e.g., returns a NaN, it triggers a trap event.
In cases where a VLIW requires 512 clocks or more due to a set stall signal, an abort and trap event will be triggered. It's worth noting that this threshold for clock cycles can be adjusted if needed.
\Cref{fig:stall} shows how the stall signal support is implemented.
As long as the data is not available, the stall is held and the same VLIW stalls.
Once the data is there, the VLIWs continue execution on the reconfigurable fabric.
If the stall limit is passed, the trap event will be raised to signal the exception.

The second group of traps is user-specific. Users can initiate these traps via a command within the VLIW based on the logic of the algorithm accelerated, e.g., the output is negative where output should be positive only. To achieve this, 3 bits are included in the VLIW which determines the specific trap value.

\section{Benchmarking Applications}
To show the benefit of the dynamic execution support for reconfigurable processors, we develop accelerators and SIs for four different applications.
Then we highlight the need for dynamic execution within each SI and how does it benefit from it.
\label{sec:bench}
\subsection{SIFT}
David G. Lowe introduced the Scale-Invariant Feature Transform (SIFT) algorithm for image recognition, aiming to enhance various applications such as tracking, object detection, 3D image matching, and mobile robots \cite{lowe2004distinctive}. 
The SIFT algorithm has different steps. 
One of these steps is the SIFT-match step.
In this step, the features extracted from the object are compared to a reference features value. 
Based on the Euclidean distance between the extracted and reference values, the decision on whether objects match or not is taken.
This comparison follows \cref{eq:sift}. $a$ is the reference object, $b$ is the detected object, $d(a,b)$ is the euclidean distance between both objects, $a_i$ is the i-th feature of object $a$, $b_i$ is the i-th feature of object $b$, and $n$ is the total number of features. 

\begin{equation}
    \label{eq:sift}
    d(a,b) = \sum^n_i(a_i-b_i)^2
\end{equation}

We develop the SIFT-match-SI, a Special Instruction specifically designed to perform the object recognition matching kernel for the SIFT algorithm. In this process, the typical operation involves subtracting two single-precision values and then squaring the result. This operation is performed for each of the values within the feature vectors. 
After executing the Siftmatch-SI, the sum of all squared results is presented to the Integer Pipeline as a single numerical value.
To expedite the computation, it is divided into four accelerators to parallelize the execution as shown.
Each accelerator first performs the subtraction of the two values for the features and then squares the output to match the equation.

\begin{figure}[htbp]
    \centering
    \includegraphics[clip=true, width=1\linewidth,page=1]{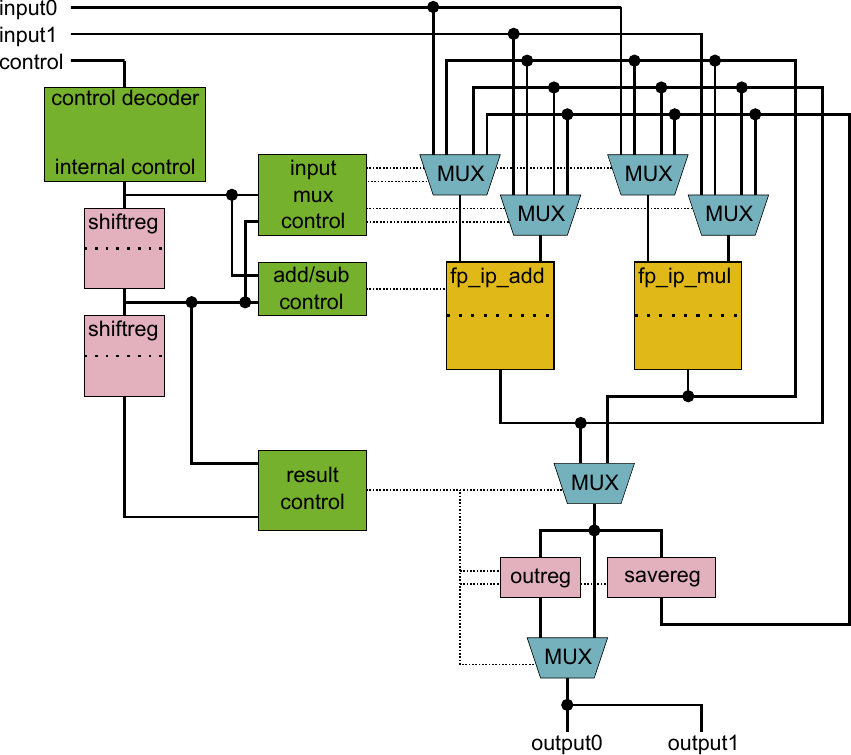}
    \caption{Accelerators design executing the SIFT-match SI}
    \label{fig:sift_fmav}
\end{figure}

\Cref{fig:sift_fmav} shows the design of the accelerator.
It contains two floating point units.
One is dedicated for addition and the other is dedicated for multiplication.
The control units and the muxes control which unit is enabled an which input is given either from the accelerator input or from the data saved.
The registers are used to provide the correct inputs, as well to save the accumulated value till the calculation within the accelerator is finalized.

\subsubsection*{Benefit to SIFT from Dynamic Execution}
The features vector can have arbitrary length, moreover, based on different constraints on the system a reduced version with a lower number of features might be executed.
Having dynamic execution, where the number of features can be set dynamically at runtime enables this. 
Otherwise, the SI will only be able to support a fixed number of features.

\subsection{SWE}

The Shallow Water Equations are a system of hyperbolic differential equations that describe the behavior of incompressible fluids, such as water, in scenarios where the horizontal dimensions are significantly smaller than the depth. These equations are derived from the Navier-Stokes equations and are typically applied to model fluid flow in 2D domains divided into cells using a consistent rectangular grid. To simulate the flow over discrete time intervals, Riemann problems are solved at the cell edges. The state of the simulation grid in the subsequent time step is then calculated using the solutions obtained from these edge-local Riemann problems \cite{breuer2012teaching}.

The Riemann equation involved in this process is quite lengthy and complex, making it an excellent candidate for testing the dynamic control. The initial solver used for this purpose was the FWave solver~\cite{breuer2012teaching}, which provides a straightforward Riemann solution with reasonable performance. However, it has a notable limitation—it can only model bodies of water and cannot represent features like shorelines because it lacks the capability to simulate the wetting or drying of cells.
Therefore, the HLLE solver which is more complex but also more general is also needed~\cite{breuer2012teaching}.

We build the SWE-SI which is able of performing the calculations from both solvers.
For this, we develop four different accelerators.
One accelerator performs the addition, multiplication, and subtraction of floating points (very similar to the accelerator from \cref{fig:sift_fmav}), another performing floating point division, a third performs square root calculation, and one does utility services like calculating max and min, and so on.
For the final design at the reconfigurable fabric, we use two FMAV accelerators to deal with the several additions, multiplications, and substractions. 
Then we have one accelerator from each type. to do the different calculations.

\begin{figure}[htbp]
    \centering
    \includegraphics[clip=true, width=1\linewidth,page=1]{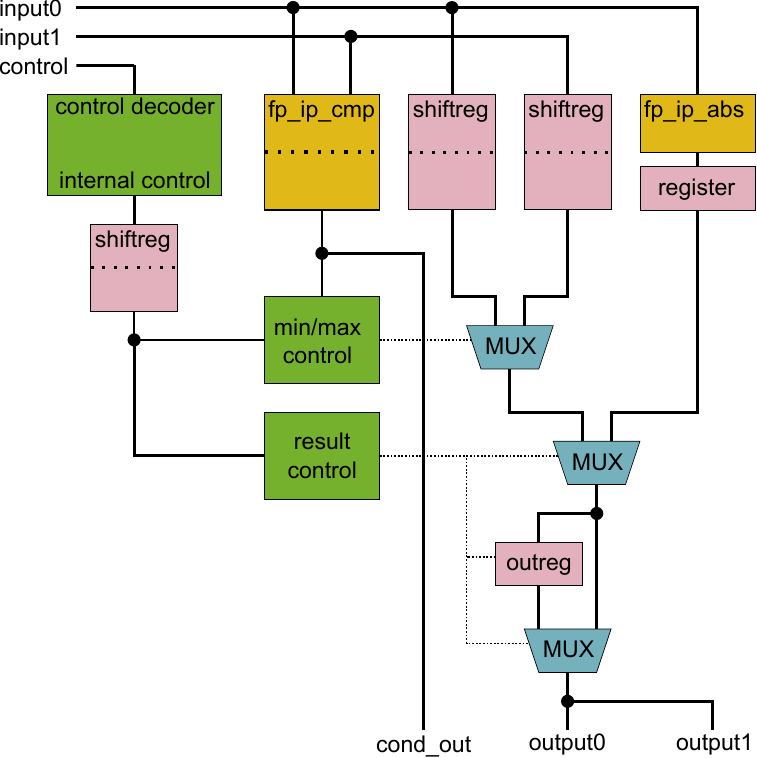}
    \caption{Accelerator designed for utility acceleration for the SWE-SI.}
    \label{fig:swe_util}
    \vspace{-2em}
\end{figure}

\Cref{fig:swe_util} shows the design of the utility accelerator as an example of the accelerators developed.
We have one floating point comparator unit and one unit that calculate absolute value of floating point numbers.
Inputs are given to the units directly from the inputs of the accelerators.
The control units and muxes decide which output from both units.

\subsection*{Benefit to SWE from Dynamic Execution}
For SWE as we have two solvers, we need to dynamically switch between them based on the surrounding environment. 
If we have wetting of dry surroundings, we can use the HLLE microcode, otherwise, we use the FWAVE microcode.
Therefore, dynamic execution helps us to jump between both microcodes at runtime.

\subsection{CNN}

\begin{figure}[htbp]
    \centering
    \includegraphics[clip=true, width=1\linewidth,page=1]{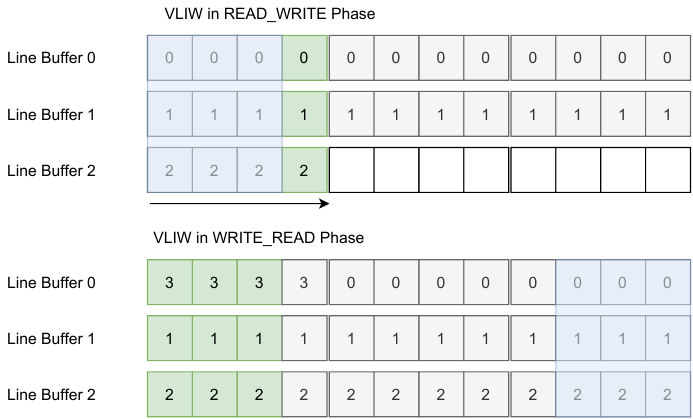}
    \caption{Line buffers used for the CNN-MAC accelerator. Streaming data to the buffers significantly reduces the bottleneck for both writing and reading the data.}
    \label{fig:cnn_mac}
    \vspace{-1em}
\end{figure}

The utilization of Convolutional Neural Networks (CNNs) presents a significant challenge due to their demanding computational and storage requirements~\cite{NEURIPS2021_20568692,NIPS2012_c399862d,russakovsky2015imagenet}. 
In order to accelerate CNNs with reconfigurable processors we design two accelerators.
The first accelerator is the CNN-MAC accelerator.
This is a multiply accumulate unit capable of accelerating $3\times 3$ matrices needed for CNN calculation.
CNNs suffer from a memory bottleneck. 
For this purpose, each CNN-MAC accelerator has 3 line buffers implemented in block RAMs as \cref{fig:cnn_mac} shows.
Data is continuously streamed to the buffers so they are ready once needed.

The second accelerator we designed is CNN-SUM.
This accelerator performs the activation, pooling and quantization steps for the CNN.
Similar to the case with CNN-MAC, CNN-SUM also has line buffers where data are streamed to resolve the memory bottleneck.
The final setup of the reconfigurable fabric to accelerate CNNs contains two CNN-SUM and two CNN-MAC atoms are used to have as much parallel execution as possible.

\subsubsection*{Benefit to CNN from Dynamic Execution}
\begin{figure}[htbp]
    \centering
    \includegraphics[clip=true, width=0.5\linewidth,page=1]{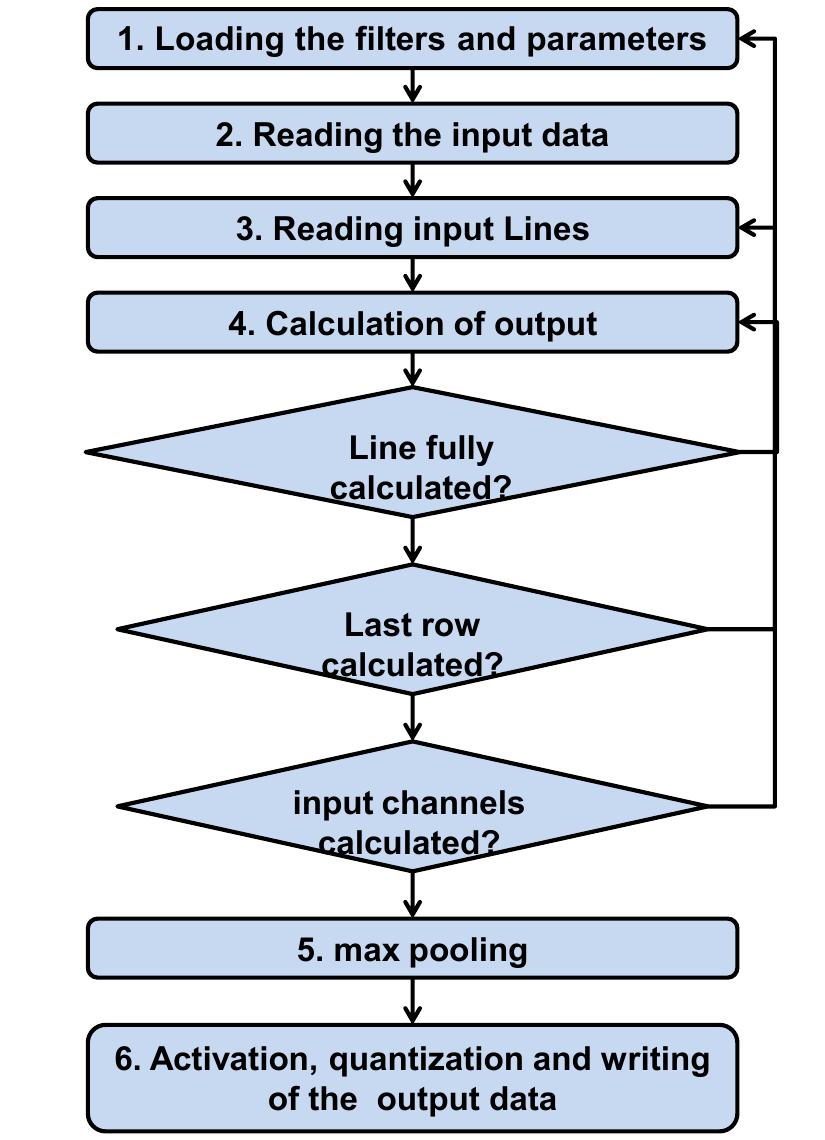}
    \caption{Steps performed for the CNN-acceleration. The SI contains three nested loops which require our dynamic execution. Other SIs are not shown for brevity.}
    \label{fig:cnn_si}
    \vspace{-2em}
\end{figure}

To build a generic accelerator for CNNs, it needs to support an arbitrary number of channels.
Without the dynamic execution, our reconfigurable processor would not have the capability to execute an arbitrary number of channels.
It would have needed instead to design several SI microcodes to support each and every channel size.
\Cref{fig:cnn_si} shows the steps of the developed CNN SI.
It contains three nested loops wich show the need for the dynamic execution, specially to deal with the arbitrary unumber of channels.

\subsection{SHA-3}

SHA-3 stands for Secure Hash Algorithm 3, and it's a family of cryptographic hash functions that have been defined by the U.S. National Institute of Standards and Technology (NIST) and are proven to be post-quantum secure~\cite{SHA-3}.
To process queries of varying lengths, hash functions typically comprise three components. First, a padding function is employed to expand the input data to an integer multiple of the block length required by the hash function. This enables the input to be divided into several blocks, each of equal size.

Following this, a compression function (or, in the case of SHA-3, a permutation) is constructed. This function sequentially combines the blocks with the output of the permutation and further processes them. While many widely used hash functions follow the Merkle-Damgård construction, SHA-3 adopts the sponge construction.
In this process, the input data is divided into multiple equally sized blocks through the use of padding. These blocks are then successively combined using the sponge construction, resulting in a 1600-bit bit vector. The final hash is extracted from this wide-bit vector~\cite{SHA-3}.

\begin{figure}[htbp]
    \centering
    \includegraphics[clip=true, width=1\linewidth,page=1]{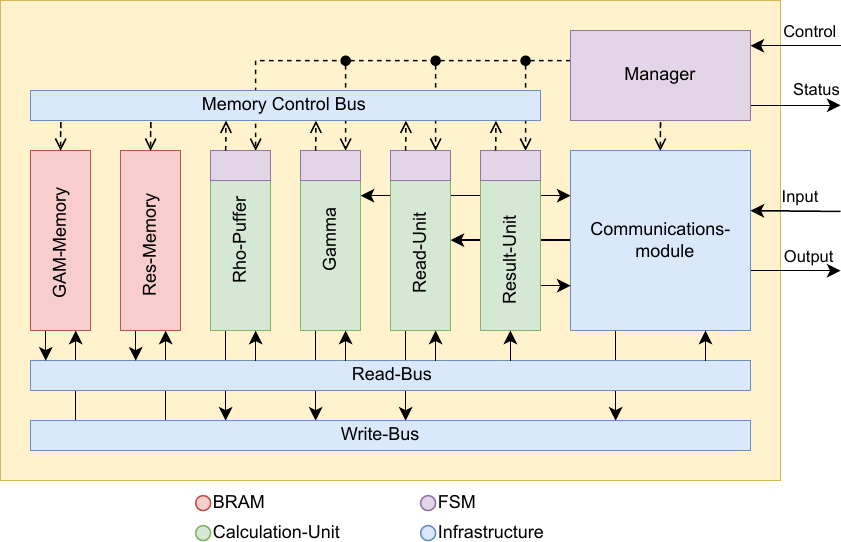}
    \caption{The internal structure of the SHA-Comp accelerator}
    \label{fig:sha_comp}
    \vspace{-1em}
\end{figure}

To accelerate SHA-3 we design two accelerators.
The first accelerator is SHA-Buff.
It is used to fetch data from memory.
As for SHA, the data can be quite large and of any arbitrary length it can easily become a bottleneck.
Therefore, we need to stream data to the BRAMs to resolve the bottleneck.
The second accelerator is SHA-Comp.
It does all the computation needed for SHA.
\Cref{fig:sha_comp} shows the internal design of the accelerator.
It contains two RAM components the GAM-memory which stores the output of the gamma step of the application.
The second RAM component is the Res-memory which stores the output of the whole algorithm, it also stores the intermediate outputs till the final output is reached.
It has four calculation units, Gamma for the gamma part of the application, read and result units which get the input and give out the output and the Rho buffer which shifts the data to perform the permutation step.

\begin{figure}[htbp]
    \centering
    \includegraphics[clip=true, width=1\linewidth,page=1]{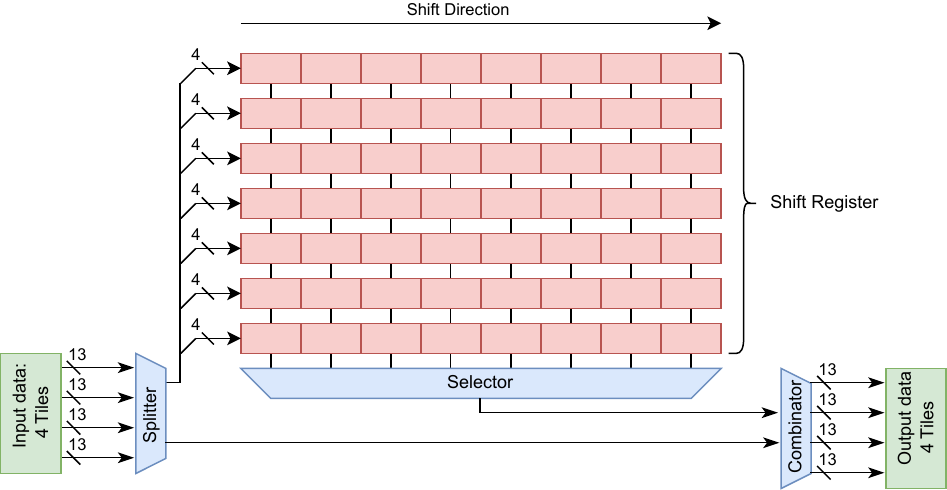}
    \caption{Structure of the Rho buffer}
    \label{fig:sha_rho}
    \vspace{-1em}
\end{figure}

The architecture of the rho-buffer is shown in \cref{fig:sha_rho}.
It gets four inputs, they get splitted using the splitter and then fed to seven shift registers.
After shifting the data, using a selector, the data is selected and then combined to get the outputs.

The final structure of the reconfigurable fabric uses two SHA-Buff accelerators and two SHA-Comp accelerators. 
The data streamed from each SHA-Buff accelerator is then fed directly to one of the SHA-Comp accelerators.
Then each of the SHA-Comp accelerators performs the post-quantum-secure algorithm on part of the data in parallel to have maximum acceleration.

\subsubsection*{Benefit to SHA from Dynamic Execution}
Hash algorithms have to be able to support several input data sizes.
To achieve this, it needs a dynamic loop that can adapt to the size of the data at runtime.
This is not simple, as for example, the amount of padding needed will change based on the exact size of the data.
Therefore, the dynamic execution allows the SHA-3 SI to be able to work with data of any arbitrary size instead of fixing on one size.
%%%%%%%%%%%%%%%%%%%%%%%%%%%%%%%%%%%%%%%%%%%%%%%%%%%%%%%%%%%%%%%%%%%%%%%%%%%%%%%%%%%%%
\section{Evaluation}
\label{sec:results}
\subsection{Experimental Setup}
All tests were conducted on a Xilinx VC707 development board, which features a Virtex-7 XC7VX485T-2FFG1761C FPGA.
The CPU core used is a Leon3 core~\cite{leon3}.
The reconfigurable fabric is implemented to contain 5 accelerator slots.
Each slot has 1600 LUTs, 3200 Flip-Flops, 10 Brams, and 20 DSPs.
The benchmark applications were loaded onto the board and executed as "bare-metal" applications and were compiled using the Gaisler BCC version 4.4.2 compiler~\cite{leon3}, which we modified to support using the SIs.

To configure the FPGA, bitstreams were generated using Vivado. To assess FPGA utilization and timing performance, data was collected through the "Report Timing Summary" and "Report Utilization" features provided by Vivado.

\subsection{Resource Usage}

To introduce the dynamic execution, a hardware extension is developed.
Moreover, while all accelerators have an upper bound of resources, each accelerator designed for the different SIs has its own resource utilization.
The resource utilization of the different components of our system are shown in \cref{tab:Area}.
It can be seen that in comparison to the Leon3 core, the dynamic execution hardware is $7\times$ smaller which is an acceptable overhead.

As for the accelerators, the CNN accelerators are the most resource-intensive with CNN-MAC hitting the upper limit of available LUTs.
The SHA-Comp accelerator has also a similar resource utilization.
Moreover, for both CNN and SHA accelerators having BRAM within the accelerator is very crucial as they use several of them.
This is especially relevant for the SHA-Buff accelerator which is basically a RAM buffering needed data.
In contrast, for SIFT and SWE, the accelerators are less resource hungry.
However, as both basically execute mathematical equations, the usage of DSP is required.

\begin{table}[htb]
  \centering
  \caption{Resources needed for each component}
  \label{tab:Area}
  \begin{tabular}{c c c c c}
    \hline
    Component & DSP & LUT & BRAM & Flip-Flop\\
    \hline
    Leon3 & 0 & 14013 & 18 & 6924 \\
    \hline
    Dynamic Execution & 0 & 2100 & 0 & 682 \\
    \hline
    SIFT-FMAV & 4 & 627 & 0 & 277\\
    \hline
    SWE-SQRT & 0 & 526 & 0 & `85\\
    SWE-UTIL & 0 & 179 & 0 & 206\\
    SWE-FMAV & 4 & 627 & 0 & 277\\
    SWE-DIV  & 0 & 857 & 0 & 258\\
    \hline
    SHA-Comp & 0 & 1205 & 2 & 628\\
    SHA-Buff & 0 & 0 & 4 & 0\\
    \hline
    CNN-SUM & 5 & 1333 & 8 & 249\\
    CNN-MAC & 9 & 1600 & 10 & 3200 \\
    \hline
  \end{tabular}
  \vspace{-2em}
\end{table}
\subsection{Timing Improvement}
\begin{figure}[htbp]
    \centering
    \includegraphics[clip=true, width=1\linewidth,page=1]{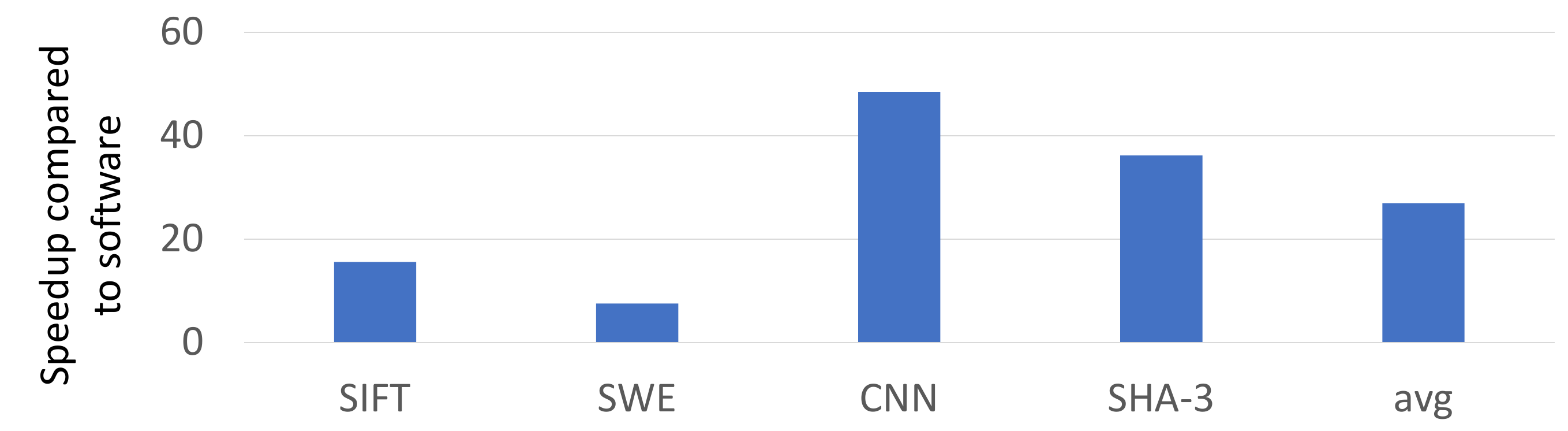}
    \caption{Execution time improvement of the benchmarking applications. The memory intensive applications (SHA-3 and CNN have remarkably higher timing improvement. On average we have $27\times$ improvement of execution time.}
    \label{fig:time_chart}
    \vspace{-2em}
\end{figure}

The goal of developing the dynamic execution for the SIs executed on the reconfigurable fabric is to improve the performance.
\Cref{fig:time_chart} shows the timing improvement for each of the accelerated algorithms.
The numbers are taken by running the same application on the same data-set once using the hardware acceleration and another using pure software.
The same experiment is run in a loop then for 1000 times then averaged to make sure that any noise in the performance gets eliminated.
CNN and SHA-3 get the highest speedups. 
This is expected as both applications are memory bound.
Using the buffering of the data to the accelerators to use them right away solves the problem.
Having the dynamic execution eases this by allowing to process a dynamic number of channels for CNN or data with arbitrary size for for SHA-3.
SIFT and SWE have speedups of $14\times$ and $7\times$ respectively which is still significant.
%%%%%%%%%%%%%%%%%%%%%%%%%%%%%%%%%%%%%%%%%%%%%%%%%%%%%%%%%%%%%%%%%%%%%%%%%%%%%%%%%%%%
\section{Conclusions}
\label{sec:conclusion}
In conclusion, this work introduces dynamic execution for reconfigurable processors. 
We add support for different jump sub-instructions, stall the processor, and several exception classes.
We show that the dynamic execution is very useful in accelerating a spectrum of applications ranging from object detection to security applications.
We create special instructions for four different applications and develop the accelerators needed for each of the applications.
Our results show that using our SIs and accelerators we have a relatively low overhead of 6\% and we are able to reach an execution time improvement up to $42\times$ and on average to $27\times$.

\bibliographystyle{IEEEtran}
\bibliography{sidechannel}

@InProceedings{iccad21,
  Title = {{LoopBreaker}: Disabling Interconnects to Mitigate Voltage-Based Attacks in Multi-Tenant {FPGAs}},
  Author = {Nassar, Hassan and AlZughbi, Hanna and Gnad, Dennis and Bauer, Lars and Tahoori, Mehdi and Henkel, J\"{o}rg},
  Booktitle = {ICCAD},
  Year = {2021}
}

@INPROCEEDINGS{muteco21,
  author={Jordan, Michael Guilherme and Korol, Guilherme and Rutzig, Mateus Beck and Beck, Antonio Carlos Schneider},
  booktitle={SBCCI}, 
  title={{MUTECO: A Framework for Collaborative Allocation in CPU-FPGA Multi-tenant Environments}}, 
  year={2021}}

@INPROCEEDINGS{NGCAS2017DPRNOC,
  author={Hassan, Amr and Mostafa, Hassan and Fahmy, Hossam A. H. and Ismail, Yehea},
  booktitle={NGCAS}, 
  title={Exploiting the Dynamic Partial Reconfiguration on {NoC}-Based {FPGA}}, 
  year={2017}}

@INPROCEEDINGS{icm22,
  author={Elsaid, Kareem and Safar, Mona and El-Kharashi, M. Watheq},
  booktitle={ICM}, 
  title={{Optimized FPGA Architecture for Machine Learning Applications using Posit Multipliers}}, 
  year={2022}}

@Incollection{bauer_fau2022,
  title={Adaptive Application-Specific Invasive Micro-Architectures},
  author={Bauer, Lars and Becker, J{\"u}rgen and Henkel, J{\"o}rg and Lesniak, Fabian and Nassar, Hassan},
  bookTitle="Invasive Computing",
year="2022",
publisher="FAU University Pres"}

@inproceedings{bauer2009rispp,
  title={{RISPP: A run-time adaptive reconfigurable embedded processor}},
  author={Bauer, Lars and Shafique, Muhammad and Henkel, J{\"o}rg},
  booktitle={IEEE FPL},
  year={2009}
}

@inproceedings{breuer2012teaching,
  title={Teaching parallel programming models on a shallow-water code},
  author={Breuer, Alexander and Bader, Michael},
  booktitle={IEEE ISPDC},
  year={2012}
}

@article{lowe2004distinctive,
  title={Distinctive image features from scale-invariant keypoints},
  author={Lowe, David G},
  journal={International journal of computer vision},
  year={2004}
}

@misc{SHA-3,
  author = {Morris Dworkin},
  title = {SHA-3 Standard: Permutation-Based Hash and Extendable-Output Functions},
  year = {2015},
  publisher = {NIST FIPS}
}

@inproceedings{NIPS2012_c399862d,
 author = {Krizhevsky, Alex and Sutskever, Ilya and Hinton, Geoffrey E},
 booktitle = {Advances in Neural Information Processing Systems},
 editor = {F. Pereira and C.J. Burges and L. Bottou and K.Q. Weinberger},
 title = {ImageNet Classification with Deep Convolutional Neural Networks},
 year = {2012}
}

@misc{russakovsky2015imagenet,
      title={ImageNet Large Scale Visual Recognition Challenge}, 
      author={Olga Russakovsky and Jia Deng and Hao Su and Jonathan Krause and Sanjeev Satheesh and Sean Ma and Zhiheng Huang and Andrej Karpathy and Aditya Khosla and Michael Bernstein and Alexander C. Berg and Li Fei-Fei},
      year={2015},
      archivePrefix={arXiv}
}

@inproceedings{NEURIPS2021_20568692,
 author = {Dai, Zihang and Liu, Hanxiao and Le, Quoc V and Tan, Mingxing},
 booktitle = {Advances in Neural Information Processing Systems},
 editor = {M. Ranzato and A. Beygelzimer and Y. Dauphin and P.S. Liang and J. Wortman Vaughan},
 title = {CoAtNet: Marrying Convolution and Attention for All Data Sizes},
 year = {2021}
}

@manual{leon3,
title = {GRLIB VHDL IP Core Library: Configuration and Development Guide},
date = {June 2023},
version = {2023.2},
year = {2023},
organization = {Cobham Gaisler AB}
}

@InProceedings{Molen,
author="Kuzmanov, Georgi
and Gaydadjiev, Georgi
and Vassiliadis, Stamatis",
title={{The Virtex II ProTM MOLEN Processor}},
booktitle="Computer Systems: Architectures, Modeling, and Simulation",
year="2004"
}

@inproceedings{kharisma,
author = {Koenig, Ralf and Bauer, Lars and Stripf, Timo and Shafique, Muhammad and Ahmed, Waheed and Becker, Juergen and Henkel, J\"{o}rg},
title = {{KAHRISMA: A Novel Hypermorphic Reconfigurable-Instruction-Set Multi-Grained-Array Architecture}},
year = {2010},
booktitle = {DATE}
}

@manual{zynq,
title = {Zynq UltraScale+ Device Technical Reference Manual},
date = {April 2023},
version = {2.3.1},
year = {2023},
organization = {AMD}
}

@manual{stratix,
title = {Intel Stratix 10 SoC FPGA Boot User Guide},
date = {September 2023},
version = {22.4},
year = {2023},
organization = {Intel}
}

@INPROCEEDINGS{autoSI,
  author={Harbaum, Tanja and Schade, Christoph and Damschen, Marvin and Tradowsky, Carsten and Bauer, Lars and Henkel, Jörg and Becker, Jürgen},
  booktitle={IEEE SOCC}, 
  title={{Auto-SI: An adaptive reconfigurable processor with run-time loop detection and acceleration}}, 
  year={2017}}

@INPROCEEDINGS{iccad19,
  author={Damschen, Marvin and Bauer, Lars and Henkel, Jörg},
  booktitle={ICCAD}, 
  title={{WCET Guarantees for Opportunistic Runtime Reconfiguration}}, 
  year={2019}}

@inproceedings{corefab,
author = {Grudnitsky, Artjom and Bauer, Lars and Henkel, J\"{o}rg},
title = {{COREFAB: Concurrent Reconfigurable Fabric Utilization in Heterogeneous Multi-Core Systems}},
year = {2014},
booktitle = {CASES}
}

\vfill

\end{document}